\RequirePackage{luatex85}
\documentclass[reprint,twocolumn,superscriptaddress,showpacs,nofootinbib,notitlepage,preprintnumbers,aps,prd,floatfix]{revtex4-2}
\usepackage{qcircuit}
\usepackage{graphicx}
\usepackage{amsmath}
\usepackage{standalone}
\usepackage{relsize}
\usepackage{multirow}
\usepackage{pbox}
\usepackage{booktabs}

\usepackage{ifthen}
\usepackage{xparse}
\usepackage{units}
\usepackage{mathtools}
\usepackage{pgffor}
\usepackage{subcaption}
\usepackage{hyperref}
\usepackage[nomessages]{fp}

\DeclarePairedDelimiter{\ketc}{\vert}{\rangle}
\DeclarePairedDelimiterX{\braketctwo}[2]{\langle}{\rangle}{%
#1\,\delimsize\vert\,\mathopen{}#2}
\DeclarePairedDelimiterX{\braketcthree}[3]{\langle}{\rangle}{#1\,\delimsize\vert\,\mathopen{}#2\,\delimsize\vert\,\mathopen{}#3}
\DeclarePairedDelimiterX{\ketbrac}[2]{\lvert}{\rvert}{\,\mathopen{}#1\,\delimsize\rangle\delimsize\langle\,\mathopen{}#2\,}
\DeclarePairedDelimiter{\parc}{(}{)}

\DeclareMathOperator{\Add}{Add}
\DeclareMathOperator{\FMA}{FMA}
\let\Reset\relax
\DeclareMathOperator{\Reset}{Reset}
\DeclareMathOperator{\Copy}{Copy}
\DeclareMathOperator{\Shift}{Shift}
\DeclareMathOperator{\Negate}{Negate}
\DeclareMathOperator{\Mult}{Mult}
\DeclareMathOperator{\Swap}{Swap}
\DeclareMathOperator{\ZeroExp}{ZeroExp}
\DeclareMathOperator{\Recip}{Recip}

\newcommand*{\ket}[1]{\ketc*{#1}}

\NewDocumentCommand{\braket}{m g m}{\ifthenelse{\equal{#2}{}}{\braketctwo*{#1}{#2}}{\braketcthree*{#1}{#2}{#3}}}
\NewDocumentCommand{\deriv}{g m}{\frac{\mathrm{d} #1}{\mathrm{d} #2}}
\newcommand*{\Par}[1]{\parc*{#1}}
\newcommand*{\LOp}[1]{\ifcat\noexpand#1\relax\bm{#1}\else\mathbf{#1}\fi}
\newcommand*{\T}[0]{\mathrm{T}}
\renewcommand*{\vec}[1]{\LOp{#1}}

\newcommand*{\Op}[1]{\hat{#1}}

\newcommand*{\Wi}[1]{\tfrac{1}{#1}}
\newcommand*{\Wirt}[1]{\Wi{\sqrt{#1}}}

\newcommand*{\vWirt}[1]{\vphantom{\Wirt{#1}}}

\newcommand*{\cb}[0]{\mathrm{c.b.}}

\usepackage{algpseudocode}
\usepackage{float}
\algnewcommand\algorithmicregister{\textbf{Register:}}
\algnewcommand\algorithmicregisters{\textbf{Registers:}}
\algnewcommand\Register{\item[\algorithmicregister]}
\algnewcommand\Registers{\item[\algorithmicregisters]}
\algnewcommand\algorithmiclet{\textbf{Let}}
\algnewcommand\Let{\item[\algorithmiclet]}
\usepackage{newfloat}
\DeclareFloatingEnvironment[name=ALG.]{algorithm}
\usepackage{tikz}
\usepackage{pgfplots}
\usepackage{tikzscale}
\usepackage{pgffor}

\usetikzlibrary{arrows.meta}
\usetikzlibrary{backgrounds}
\usetikzlibrary{patterns}
\usetikzlibrary{patterns.meta}
\usepgfplotslibrary{patchplots}
\usepgfplotslibrary{fillbetween}
\usepgfplotslibrary{statistics}
\usepgfplotslibrary{colormaps}
\pgfplotsset{compat=1.18}
\pgfplotsset{%
    layers/standard/.define layer set={%
        background,axis background,axis grid,axis ticks,axis lines,axis tick labels,pre main,main,axis descriptions,axis foreground%
    }{
        grid style={/pgfplots/on layer=axis grid},%
        tick style={/pgfplots/on layer=axis ticks},%
        axis line style={/pgfplots/on layer=axis lines},%
        label style={/pgfplots/on layer=axis descriptions},%
        legend style={/pgfplots/on layer=axis descriptions},%
        title style={/pgfplots/on layer=axis descriptions},%
        colorbar style={/pgfplots/on layer=axis descriptions},%
        ticklabel style={/pgfplots/on layer=axis tick labels},%
        axis background@ style={/pgfplots/on layer=axis background},%
        3d box foreground style={/pgfplots/on layer=axis foreground},%
    },
}

\usepackage{alphalph}

\usepgfplotslibrary{external}
\usepgfplotslibrary{groupplots}
\AtBeginEnvironment{figure}{\setlength{\intextsep}{0}}
\newcommand*{\affilA}{Center for Biomedical Imaging, Department of Radiology, New York University Grossman School of Medicine, New York, New York, 10016, USA}
\newcommand*{\affilB}{Superconducting and Quantum Materials System Center (SQMS), Batavia, Illinois, 60510, USA}
\newcommand*{\affilC}{Fermi National Accelerator Laboratory, Batavia, Illinois, 60510, USA}
\newcommand*{\affilD}{NASA Ames Research Center, Moffett Field, California 94035, USA}

\begin{document}
\preprint{FERMILAB-PUB-25-0270-SQMS}

\title{Efficient Floating-Point Arithmetic on Fault-Tolerant Quantum Computers}
\author{Jos{\'e} E. Cruz Serrall{\'e}s}
\email{Jose.CruzSerralles@nyulangone.org}
\affiliation{\affilA}
\author{Oluwadara Ogunkoya}
\email{ogunkoya@fnal.gov}
\affiliation{\affilB}
\affiliation{\affilC}
\author{Do{\~g}a Murat K{\"u}rk{\c{c}}{\"u}o{\~g}lu}
\email{dogak@fnal.gov}
\affiliation{\affilB}
\affiliation{\affilC}
\author{Nicholas Bornman}
\affiliation{\affilB}
\affiliation{\affilC}
\author{Norm M. Tubman}
\affiliation{\affilB}
\affiliation{\affilD}
\author{Anna Grassellino}
\affiliation{\affilB}
\affiliation{\affilD}
\author{Silvia Zorzetti}
\affiliation{\affilB}
\affiliation{\affilC}
\author{Riccardo Lattanzi}
\affiliation{\affilA}
\date{\today}
\begin{abstract}
	We propose a novel floating-point encoding scheme that builds on prior work involving fixed-point encodings. We encode floating-point numbers using Two's Complement fixed-point mantissas and Two's Complement integral exponents. We used our proposed approach to develop quantum algorithms for fundamental arithmetic operations, such as bit-shifting, reciprocation, multiplication, and addition. We prototyped and investigated the performance of the floating-point encoding scheme on quantum computer simulations by performing reciprocation on randomly drawn inputs and by solving first-order ordinary differential equations, while varying the number of qubits in the encoding. We observed rapid convergence to the exact solutions as we increased the number of qubits and a significant reduction in the number of ancilla qubits required for reciprocation when compared with similar approaches.
\end{abstract}
\maketitle

\section{Introduction}
IEEE-754 floating-point arithmetic is the foundation of numerical computing, ensuring accuracy, consistency, and efficiency across a wide range of classical CPU architectures and integrated circuits \cite{ieee754}. Whether in high-performance computing, mobile devices, or specialized hardware like graphics processing units (GPUs) and application-specific integrated circuits (ASICs), the IEEE-754 standard plays a crucial role in enabling reliable floating-point computations universally. Floating-point numbers offer fixed relative precision \cite{goldberg1991floatintro}, as opposed to fixed-point numbers that offer fixed absolute precision \cite{yates2009fixedintro}. By adjusting the exponent and mantissa within corresponding limits, floating-point numbers allow for a wider range of values and precisions than fixed-point numbers \cite{goldberg1991floatintro}. The latter are represented using integer primitives with an implicit decimal or binary point, so fixed-point arithmetic is often used in applications where deterministic precision and lower computational overhead are necessary, such as digital signal processing and financial calculations \cite{yates2009fixedintro}. However, most classical computing architectures do not provide native hardware support for fixed-point arithmetic. Instead, fixed-point operations are typically emulated using integer arithmetic, requiring additional software-based scaling and rounding techniques to achieve the desired level of numerical accuracy \cite{yates2009fixedintro}. 

The challenge of performing arithmetic operations on quantum computers lies in the precise encoding of numerical data into qubits, along with the design of optimized quantum circuits capable of efficiently executing addition, multiplication, division, and other mathematical operations. Researchers have investigated various methods for implementing both fixed-point and floating-point arithmetic \cite{zanger2021,nachtigal2011,wiebe2013,peng2014,jain2015,haener2018,sanada2019,zhang2020,rogers2020,gayathri2021a,bhaskar2015quantum,cao2013quantum,gayathri2021b,kumar2022,seidel2022,zhao2022,steijl2024}. These approaches were designed to facilitate efficient numerical computations, which are crucial for numerous applications. However, despite the development of theoretical frameworks and preliminary implementations, the majority of these methods have not been thoroughly explored nor widely adopted. A significant number of these approaches rely on brute-force techniques, using hardware description languages (HDLs) such as Verilog or VHDL to construct circuit descriptions \cite{haener2018, das2019optimizing}. While HDLs provide a straightforward way to model quantum arithmetic, these languages were designed with classical logical synthesis in mind, requiring the use of irreversible operations, such as logical \texttt{AND}, that have no direct unitary analogue on quantum computers. As a result, implementing a single irreversible operation on quantum computers would require its own set of ancillas, which quickly becomes impractical as the complexity of the quantum circuit increases, requiring a large number of irreversible operations.

In addition to techniques based on HDLs, other approaches for floating-point arithmetic include Quantum Fourier Transform (QFT) and Clifford+T gate-based arithmetic design. A comprehensive review of previous work based on QFT can be found in \cite{ruiz2017quantum,atchade2023efficient}. QFT-based approaches utilize Hadamard gates and controlled rotation gates, which include both Clifford and non-Clifford gates. Much like how the Discrete Fourier Transform diagonalizes the convolution of two sequences into multiplication in the frequency domain, the Quantum Fourier Transform operates instead over the binary states of qubits and allows one to perform arithmetic such as addition and multiplication by diagonalizing the respective operations \cite{serralles2024quantum}. In contrast, the Clifford+T-based arithmetic design relies on Clifford gates along with $T$ gates, which are used to decompose the Toffoli gate into a composition of one $H$ gate, seven $T$ gates, and six $\mathrm{CNOT}$ operations \cite{amy2013meet}.

The primary goal of this work is  to develop a floating-point arithmetic framework that minimizes the number of required ancilla qubits. To achieve this, we propose a novel QFT-based approach for efficient floating-point arithmetic that does not rely on information encoded in the wavefunction coefficients and reuses ancilla qubits by exploiting the structure of the encoding. Namely, our method strictly utilizes information embedded in the non-zero states associated with the binary strings representing a given number, whereas other approaches for quantum computation, such as amplitude encoding, phase encoding, and hybrid methods, store information in the complex weights of the wavefunction itself and differ fundamentally in their algorithmic design \cite{wiebe2013,lau2016,weigold2021a,weigold2021b,korzekwa2022,gonzalezconde2024}.

\section{Technical Background}
\begin{table*}[tbph!]
\centering
\begin{subtable}{0.49\linewidth}
\begin{tabular}{c c c c}
	\toprule
	Type              & $x$                   & $y$                   & $x-y$                   \\
	\midrule
	Analytical        & $\nicefrac{\pi}{100}$ & $\nicefrac{\pi}{128}$ & $\nicefrac{7\pi}{3200}$ \\
	64-bit floating   & $0.03142$             & $0.02454$             & $0.006872             $ \\
	16-bit fixed      & $0.03125$             & $0.02344$             & $0.007813             $ \\
	16-bit floating   & $0.03140$             & $0.02454$             & $0.006866             $ \\
	Fixed  rel. error & 0.528\%               & 4.51\%                & 13.7\%                  \\
	Float. rel. error & 0.0425\%              & 0.0308\%              & 0.084\%                 \\
	\bottomrule
\end{tabular}
\caption{Subtraction}
\label{tab:arith:add}
\end{subtable}
\begin{subtable}{0.49\linewidth}
\begin{tabular}{c c c c}
	\toprule
	Type              & $x$        & $y$                 & $x \times y$          \\
	\midrule
	Analytical        & $\pi$      & $\nicefrac{1}{100}$ & $\nicefrac{\pi}{100}$ \\
	64-bit floating   & $3.141593$ & $0.01000$           & $0.031416           $ \\
	16-bit fixed      & $3.140625$ & $0.01172$           & $0.035156           $ \\
	16-bit floating   & $3.140625$ & $0.010002$          & $0.031403           $ \\
	Fixed  rel. error & 0.03080\%  & 17.19\%             & 13.7\%                \\
	Float. rel. error & 0.03080\%  & 0.02136\%           & 0.0425\%              \\
	\bottomrule
\end{tabular}
\caption{Multiplication}
\label{tab:arith:mult}
\end{subtable}
\caption{Summary of classical arithmetic examples.}
\label{tab:arith}

\end{table*}

\begin{figure*}
	\centering
	\begin{subfigure}{\linewidth}
		\centering
		\caption{Canonical IEEE-754 representation.}
		\label{fig:floating:IEEE}
		\includegraphics[]{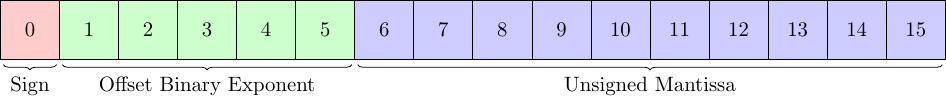}
	\end{subfigure} \\
	\begin{subfigure}{\linewidth}
		\centering
		\caption{Proposed Two's Complement representation.}
		\label{fig:floating:TC}
		\includegraphics[]{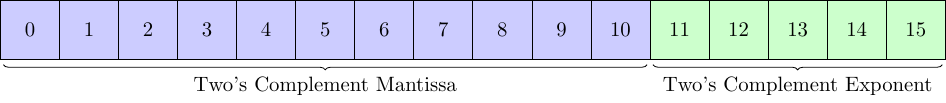}
	\end{subfigure}
	\caption{Comparison between canonical IEEE-754 encoding (\protect\subref{fig:floating:IEEE}) and floating-point encoding in this work (\protect\subref{fig:floating:TC}).}
	\label{fig:floating}
\end{figure*}

\subsection{Classical Binary Fixed- and Floating-Point Arithmetic}
Binary fixed-point and floating-point arithmetic are nearly identical in the sense that both deal with a significand that is scaled by a power-of-two exponent. In the fixed-point case, the exponent is fixed and implicit, whereas in the floating-point case, the exponent is allowed to vary and is expressed using a set of bits contained within the number, allowing one to represent a much larger set of numbers at the expense of increased algorithmic complexity when performing basic arithmetic operations. The two arithmetic operations that we focus on are addition and multiplication, as with these two operations one can implement division, exponentiation, and so on. We illustrate these differences with the following examples.

%Fixed-Point Subtraction                                        Floating-Point Subtraction                                        
%┌────────────────┬──────────────┬─────────────┬──────────────┐ ┌────────────────┬───────────────┬───────────────┬───────────────┐
%│                │            x │           y │          x-y │ │                │             x │             y │           x-y │
%├────────────────┼──────────────┼─────────────┼──────────────┤ ├────────────────┼───────────────┼───────────────┼───────────────┤
%│        float64 │  0.031415927 │ 0.024543693 │ 0.0068722339 │ │        float64 │   0.031415927 │   0.024543693 │  0.0068722339 │
%│        closest │      0.03125 │   0.0234375 │    0.0078125 │ │        closest │   0.031402588 │   0.024536133 │  0.0068740845 │
%│   approximated │      0.03125 │   0.0234375 │    0.0078125 │ │   approximated │   0.031402588 │   0.024536133 │  0.0068664551 │
%│  decomposition │       8⋅2^-8 │      6⋅2^-8 │       2⋅2^-8 │ │  decomposition │ 0.502441⋅2^-4 │ 0.785156⋅2^-5 │ 0.878906⋅2^-7 │
%│ relative error │ 0.52816057 % │ 4.5070341 % │  13.682102 % │ │ relative error │ 0.042458227 % │  0.03080137 % │ 0.084089856 % │
%└────────────────┴──────────────┴─────────────┴──────────────┘ └────────────────┴───────────────┴───────────────┴───────────────┘
Consider subtracting the numbers $\nicefrac{\pi}{100} \approx 0.031415927$ and $\nicefrac{\pi}{128} \approx 0.024543693$. Using a 16-bit Two's Complement representation with fixed exponent $2^{-8}$, the two numbers would be approximated as $8 \cdot 2^{-8} = 0.03125$ and $6 \cdot 2^{-8} = 0.0234375$, respectively, with corresponding approximate relative errors of $0.528\%$ and $4.507\%$. Subtracting the two fixed-point numbers would yield $2 \cdot 2^{-8} = 0.0078125$ with a relative error of approximately 13.7\%. Repeating the same exercise with 16-bit half-precision IEEE-754 floating-point numbers, the two numbers would be approximated as $0.5024 \cdot 2^{-4}$ and $0.785 \cdot 2^{-5}$, respectively, with corresponding approximate relative errors of $0.0425\%$ and $0.0308\%$. Subtracting the two numbers would involve repeatedly dividing the mantissa of the second number by $2$ until its exponent matches the exponent of the first input, and then subtracting the mantissas. This process of multiplying or dividing by powers of two is known as variable bit shifting. After bit shifting, the second number would have value $0.3925 \cdot 2^{-4}$. As the two numbers would have the same exponent, we would then subtract the mantissas, resulting in $0.1099 \cdot 2^{-4}$. However, the mantissa must be in the range $\left[ 0.5, 1 \right)$, so we would multiply the mantissa by $2^3$ and subtract $3$ from the exponent, resulting in the final value of $0.879 \cdot 2^{-7} \approx 0.0068664551$ with a relative error of approximately 0.084\%. Table \ref{tab:arith:add} summarizes this example.

%Fixed-Point Multiplication                                   Floating-Point Multiplication
%┌────────────────┬──────────────┬────────────┬─────────────┐ ┌────────────────┬──────────────┬───────────────┬───────────────┐
%│                │            x │          y │         x*y │ │                │            x │             y │           x*y │
%├────────────────┼──────────────┼────────────┼─────────────┤ ├────────────────┼──────────────┼───────────────┼───────────────┤
%│        float64 │    3.1415927 │       0.01 │ 0.031415927 │ │        float64 │    3.1415927 │          0.01 │   0.031415927 │
%│        closest │     3.140625 │ 0.01171875 │     0.03125 │ │        closest │     3.140625 │   0.010002136 │   0.031402588 │
%│   approximated │     3.140625 │ 0.01171875 │  0.03515625 │ │   approximated │     3.140625 │   0.010002136 │   0.031402588 │
%│  decomposition │     804⋅2^-8 │     3⋅2^-8 │      9⋅2^-8 │ │  decomposition │ 0.785156⋅2^2 │ 0.640137⋅2^-6 │ 0.502441⋅2^-4 │
%│ relative error │ 0.03080137 % │  17.1875 % │ 11.905819 % │ │ relative error │ 0.03080137 % │ 0.021362305 % │ 0.042458227 % │
%└────────────────┴──────────────┴────────────┴─────────────┘ └────────────────┴──────────────┴───────────────┴───────────────┘
As a second example, consider multiplying the numbers $\pi \approx 3.1415927$ and $\nicefrac{1}{100} = 0.01$, whose product equals approximately $0.031415297$ when using 64-bit floating-point arithmetic on a classical computer. Using a 16-bit Two's Complement representation with fixed exponent $2^{-8}$, the two numbers would be approximated as $804 \cdot 2^{-8} = 3.140625$ and $3 \cdot 2^{-8} = 0.01171875$, with relative errors of $0.031\%$ and $17.19\%$, respectively. Multiplying the two fixed-point numbers would yield $9 \cdot 2^{-8} = 0.03515625$ with a relative error of approximately 17.2\%. Repeating the same operation with 16-bit half-precision IEEE-754 floating-point numbers, the two numbers would be approximated as $0.785 \cdot 2^2$ and $0.64 \cdot 2^{-6}$, respectively. Muliplying the two numbers would involve multiplying the mantissas, adding the exponents, and adjusting the exponent if the mantissa is not in the range $\left[ 0.5,1\right)$. After multiplying the mantissas and adding the exponents, we would obtain $0.502 \cdot 2^{-4} \approx 0.042458227$ with a relative error of 0.04\%. No adjusting of the mantissa and exponent was necessary as $0.502$ falls in the allowed interval. Table \ref{tab:arith:mult} summarizes this example.

In both examples, we observed that fixed-point arithmetic suffered from loss of precision when dealing with small numbers, whereas floating-point was able to represent each number with maximal precision. Fixed-point arithmetic operations required fewer steps and essentially amounted to operations on integers with an implicit exponent. Floating-point arithmetic operations, on the other hand, were algorithmically more complex but yielded considerably more precise outputs. This increased relative precision was the main motivation in pursuing floating-point arithmetic in this work.

\subsection{Quantum Fixed-Point Arithmetic}
We recently introduced a suite of quantum algorithms for performing fixed-point arithmetic--addition, multiplication, and division--within the framework of gate-level quantum computation, using a QFT-based  approach \cite{serralles2024quantum}. These algorithms leverage the encoding of quantum states as eigenstates of the discretized position operator, enabling efficient numerical operations within the quantum domain. In this representation, we denoted qubits containing a superposition of fixed-point values using bra-ket notation. We defined a quantum register $\ket{a}$ containing fixed-point values with $n$ total qubits, and $f$ qubits after the binary point as follows. 
\begin{equation} \label{eq:fixedpoint}
	\ket{a} = \ket{a_{n-1} a_{n-2} \ldots a_{f} . a_{f-1} \ldots a_{0}} \equiv \ket{\sum_{k=0}^{n-1} a_k 2^{k-f}}
\end{equation}
We referred to such a register as a $\Par{n,f}$ unsigned fixed-point register. Note that constant fixed-point numbers (including integers) are denoted  without the ket $\ket{\cdot}$. Since Two's Complement signed fixed-point arithmetic is equivalent to unsigned fixed-point arithmetic, we refer to signed fixed-point registers using the same notation, and recycle the same operations as in the unsigned case. Given a signed fixed-point register $\ket{a}$, the definition is nearly identical to that of Eq. \ref{eq:fixedpoint}, with a slight modification to include negative numbers, which is detailed in the following equation.
\begin{equation}
    \ket{a} \equiv \ket{-a_{n-1} 2^{n-1-f} + \sum_{k=0}^{n-2} a_k 2^{k-f}}
\end{equation}
The addition algorithm was designed in a general form to support both signed and unsigned integers. This circuit remains relatively compact due to its linear structure and modest gate requirements. 
We denoted the in-place addition operation involving fixed-point numbers as $\Add\Par{\cdot}$. For example,
\begin{equation}
	\Add\Par{\ket{a},c}\colon \ket{a} \rightarrow \ket{a+c}
\end{equation}
denotes the in-place addition of the fixed-point register with state $\ket{a}$ and a fixed-point constant $c$, resulting in a new state $\ket{a+c}$. Similarly,
\begin{equation}
	\Add\Par{\ket{a},\ket{b}}\colon \ket{a,b} \rightarrow \ket{a+b,b}
\end{equation}
denotes the in-place addition of fixed-point registers with states $\ket{a}$ and $\ket{b}$, ending in respective states $\ket{a+b}$ and $\ket{b}$.

For multiplication, we extended previous techniques, notably those presented in \cite{zanger2021}, by computing the full product prior to rounding, achieving an exact computation of the product when compared with classical fixed-point products. However, this exactness comes at the cost of increased resource usage, particularly in the form of additional ancilla qubits. Note that these ancillae are not wasted, as they are strategically re-used across different steps of the computation to optimize resource efficiency. The multiplication algorithm employs a fused three-register architecture, involving doubly controlled rotation gates, which considerably increases the circuit’s overall gate count compared to the addition operation.

We denoted the in-place fused multiplication-addition (FMA) of three fixed-point numbers as $\FMA\Par{\cdot}$. For example,
\begin{equation}
	\FMA\Par{\ket{a},\ket{b},\ket{c}}\colon \ket{a,b,c} \rightarrow \ket{a+b\cdot c,b,c}
\end{equation}
denotes the in-place fused multiplication-addition of fixed-point registers with states $\ket{a}$, $\ket{b}$, and $\ket{c}$, resulting in respective states $\ket{a+b \cdot c}$, $\ket{b}$, and $\ket{c}$.

To implement division, we introduced a reciprocal operation based on Newton-Raphson's root-finding method. While the resulting algorithm is inherently approximate rather than exact, it exhibits strong convergence properties, especially when initialized with a well-chosen starting point. This makes the method practically viable for a range of fixed-point applications, despite the iterative nature of the procedure. We also implemented a basic $\Negate\Par{\cdot}$ gate that computes the two’s complement -effectively the negative of each element in a superposition of integers. This was accomplished by first applying an $\Op{X}$ gate to each qubit to perform bitwise inversion, followed by an increment operation using the $\Add\Par{\cdot}$ gate to complete the negation. We include algorithmic description of the negation operation in Alg. \ref{alg:negate}.

Together, these algorithms represent a foundational step toward building scalable arithmetic components for quantum processors, particularly those that operate on fixed-point representations within the QFT paradigm. We implemented the above operations using the Clifford gates, namely basic Pauli $\Op{X}, \Op{Y}, \Op{Z}$ gates, Hadamard $\Op{H}$ gates, and $z$-rotation $\Op{R}\Par{\theta}$ gates, along with their controlled and doubly-controlled variants. These gates admit the following basic definitions.
\begin{equation}
\begin{gathered}
	\Op{X} = \begin{bmatrix} 0 & \hphantom{-}1 \\ 1 & \hphantom{-}0 \end{bmatrix}\!\!,\hspace{1em}
    \Op{Y} = \begin{bmatrix} 0 & -i \\ i & \hphantom{-}0 \end{bmatrix}\!\!,\hspace{1em}
    \Op{Z} = \begin{bmatrix} 1 & \hphantom{-}0 \\ 0 & -1 \end{bmatrix}\!\!,\hspace{1em} \\
    \Op{H} = \frac{1}{\sqrt{2}} \begin{bmatrix} 1 & \hphantom{-}1 \\ 1 & -1 \end{bmatrix}\!\!,\hspace{1em}
    \Op{R}\Par{\theta} = \begin{bmatrix} 1 & 0 \\ 0 & e^{i \theta} \end{bmatrix}
\end{gathered}
\end{equation}
We use the above-described fixed-point addition and multiplication operations extensively in our proposed implementation of floating-point arithmetic, as the primitive operations used when performing arithmetic on the mantissa and exponent parts of the floating-point numbers.

\section{Algorithm}

\subsection{Notational Conventions}
\noindent We denote the adjoint of an operator with superscripted $\dagger$. For example, $\Add^\dagger$ denotes the adjoint of the addition operator, which results in subtraction of the arguments. In the rest of this work, we make extensive use of singly and doubly controlled gates. We denote these gates by prepending $C$ to the name of the gate, as is common in the literature. For example, the Toffoli gate
\begin{equation}
	CC\Op{X} \ket{a,b,c}
\end{equation}
denotes applying an $\Op{X}$ to $\ket{c}$ gate that is doubly controlled by qubits $\ket{a}$ and $\ket{b}$. In addition, we add a line over a qubit to denote applying an $\Op{X}$ gate to the qubit, then performing a controlled gate with the qubit, and then applying another $\Op{X}$ gate to flip the qubit back to its original state. For example,
\begin{equation} \label{eq:CCX:example}
	CC\Op{X} \ket{a,\overline{b},c}
\end{equation}
denotes applying an $\Op{X}$ gate to $\ket{b}$, applying an $\Op{X}$ gate to $\ket{c}$ that is double controlled by qubits $\ket{a}$ and $\ket{b}$, and then applying an $\Op{X}$ gate to $\ket{b}$ to flip it back to its original state. This notation is meant to resemble the logical negation operation that is common in digital circuit design because this operation effectively amounts to controlling by the inverse of the corresponding qubit.

Throughout this work, we omit the name of a gate in place of its arithmetic operation, such as $\ket{a+b}$ instead of $\Add\Par{\ket{a},\ket{b}}$. We also omit the $`C$s' denoting controlled variants and instead append \textit{c.b.}, which is shorthand for \textit{controlled by}, followed by the qubits by which we control the gate. Additionally, we will denote controlling by the logical inverse using the same convention, with the line over the qubit to be inverted and controlled by. For example, the statement
\begin{equation}
	\Op{X} \ket{c}\ \cb\ \ket{a,\overline{b}}
\label{eq:CCX:example2}\end{equation}
is equivalent to \eqref{eq:CCX:example}. The quantum circuit describing \eqref{eq:CCX:example} (hence,  \eqref{eq:CCX:example2}) is the following.

\begin{equation}
\Qcircuit @C=1em @R=1em {
  a& & \qw    &\ctrl{1}&\qw     &\qw\\
  b& &\gate{X}&\ctrl{1}&\gate{X}&\qw\\
  c& & \qw    &\targ   & \qw    &\qw
 }
 \label{circ:InvCtrl} 
\end{equation}

\subsection{Resetting Ancilla Qubits} % better explain first par.
One problem when implementing floating-point operations on quantum hardware is the reliance on ancilla qubits that serve as ``scratch space'' and whose values we discard after an operation. The issue with ancilla qubits arises from the constraint that the ancillas start in the $\ket{0}$ or ground state. However, once an operation is performed, these ancilla qubits are not necessarily in the ground state and are often entangled with other qubits. For example, let us consider a three-qubit system, with the least significant (rightmost) qubit as an ancilla qubit. The system starts with an initial state
\begin{equation} \label{eq:ancexample:0}
	\ket{\psi_0} = \ket{000}\text{.}
\end{equation}
Applying a Hadamard gate to the most significant qubit would yield the following state.
\begin{equation} \label{eq:ancexample:1}
	\ket{\psi_1} = \Wirt2 \Par{ \vWirt2 \ket{0} + \ket{1} } \otimes \ket{00} = \Wirt2 \Par{ \vWirt2 \ket{000} + \ket{100} }
\end{equation}
Then, if we perform an operation (e.g., multiplication) that entangles the ancilla register with other qubit registers such that we obtain the following state:
\begin{equation} \label{eq:ancexample:2}
	\ket{\psi_2} = \Wirt2 \Par{ \vWirt2 \ket{001} + \ket{110} }
\end{equation}
The ancilla qubit is hence in a uniformly superposed state, with its value dependent on the rest of the superposition. Measuring the ancilla qubit in order to reset the qubit would collapse the superposition to one of the two states with probability \nicefrac{1}{2}, hence destroying the information stored in the entangled registers. This would defeat the purpose of using a quantum computer for basic operations, since it would prevent taking advantage of quantum parallelism when performing computations.

Following conventional approaches, there would be two alternatives to overcome this problem: to reverse the computation involving the ancilla qubits, or to simply discard the entangled ancilla qubits and use new ancilla qubits that are known to be at ground state. While reversing the computation is possible, in principle, for all possible unitary operations on a quantum computer, doing so can prove difficult if one wants to preserve the non-ancilla state after applying one of these transformations. Also, this approach would add immensely to the overall gate cost. On the other hand, having an endless pool of ancilla qubits is practical only for small, simple problems. For example, as we showed in \cite{serralles2024quantum}, each multiplication of fixed-point registers with $F$ qubits after the decimal point would require $F$ new ancilla qubits, which would be completely impractical for problems involving many multiplications.

In light of these limitations of conventional approaches, in this work, we propose an approach in which we apply a Hadamard gate to each ancilla qubit prior to measurement. This effectively creates two trajectories for storing the main qubits' information, which is useful for decoupling the ancilla qubits from the rest of the wavefunction. Note that this is similar to the separating trajectories observed when a coupled cavity-transmon system is driven by a microwave field in the presence of a dispersive interaction, with the transmon initially prepared as a state on the equator of its Bloch sphere \cite{wallraff2004strong}. Following our proposed approach, if we measure a value of $1$, we apply an $\Op{X}$ gate to force the ancilla qubit to its ground state $\ket{0}$. To see this in practice, we continue with the example of Eqs. \eqref{eq:ancexample:0} to \eqref{eq:ancexample:2}. We first apply a Hadamard gate to the ancilla qubit in eqn \eqref{eq:ancexample:2}, resulting in the following state:
\begin{multline}
	\ket{\psi_3} = \Wirt2 \ket{00} \otimes \Wirt2 \Par{ \vWirt2 \ket{0} - \ket{1} }  \\ + \Wirt2 \ket{11} \otimes \Wirt2 \Par{ \vWirt2 \ket{0} + \ket{1} }
\end{multline}
The ancilla qubit is now in a superposed state that is independent of the states in the original superposition, which allows us to factorize and rewrite the state as follows.
\begin{multline}
	\ket{\psi_3} = \Wirt2 \Par{ \vWirt2 \ket{00} + \ket{11} } \otimes \Wirt2 \ket{0} \\ - \Wirt2 \Par{ \vWirt2 \ket{00} - \ket{11} } \otimes \Wirt2 \ket{1}
\end{multline}
Therefore, no matter which value we obtain when measuring and resetting the ancilla qubit, we always preserve the original states along with the desired entangled outputs, while possibly picking up a relative phase factor of $e^{i \pi} = -1$. At this point, we would measure the ancilla qubit, resulting in a value of $\ket{0}$ with probability $\nicefrac{1}{2}$, requiring no further action, or in a value of $\ket{1}$ with probability $\nicefrac{1}{2}$, requiring the application of an $\Op{X}$ gate to set the qubit to $\ket{0}$. We would then obtain one of the following final states, with the minus sign corresponding to measuring $\ket{1}$ in the ancilla qubit.
\begin{equation}
	\ket{\psi_4} = \Wirt2 \Par{ \vWirt2 \ket{110} \pm \ket{000} }
\end{equation}
The ancilla qubit is now in the ground state, as desired, and the rest of the superposition remains intact. This process can then be generalized to any number of ancilla qubits, wherein each ancilla qubit would be reset individually, either sequentially or in parallel.

\begin{algorithm}[tb]
	\begin{algorithmic}[1]
		\Registers input/output $\ket{q} = \ket{q_{n-1}\ldots q_0}$; input $\ket{s} = \ket{s_{m-1}\ldots s_0}$;
		ancillae $\ket{a} = \ket{a_3 a_2 a_1 a_0}$.
		\State $\ket{a_3} \leftarrow \ket{q_{n-1}}$\Comment{Copy (store sign bit)}
		\State $\ket{q} \leftarrow \ket{-q}\ \cb\ \ket{a_3}$\Comment{$C\Negate$ (abs. value)}
		\For{$k \leftarrow \Par{m-2}$ \textbf{to} $0$}
			\State $d \leftarrow 2^k$
			\State $\ket{a_1} \leftarrow \Op{X} \ket{0}\ \cb\ \ket{s_k,\overline{s_{n-1}}}$ \Comment{$CC\Op{X} \circ \Reset$}
			\If{$d < n$}
				\For{$l \leftarrow 0$ \textbf{to} $\Par{n-d-1}$}
					\State $\ket{q_l,q_{l+d}} \leftarrow \ket{q_{l+d},q_l}\ \cb\ \ket{a_1}$ \Comment{$C$Swap}
				\EndFor
			\EndIf
			\For{$l \leftarrow \max\Par{n-d,0}$ \textbf{to} $n-1$}
				\State $\ket{a_0} \leftarrow \ket{q_l}$ \Comment{Copy}
				\State $\ket{q_l} \leftarrow \Op{X} \ket{q_l}\ \cb\ \ket{a_0,a_1}$ \Comment{$CC\Op{X}$}
			\EndFor
		\EndFor
		\State $\ket{a_2} \leftarrow \ket{s_{n-1}}$\Comment{Copy (store sign bit)}
		\State $\ket{s} \leftarrow \ket{-s}$\Comment{Negate}
		\For{$k \leftarrow \Par{m-1}$ \textbf{to} $0$}
			\State $d \leftarrow 2^k$
			\State $\ket{a_1} \leftarrow \Op{X} \ket{0}\ \cb\ \ket{s_k,a_2}$ \Comment{$CC\Op{X} \circ \Reset$}
			\If{$d < n$}
				\For{$l \leftarrow \Par{n-1}$ \textbf{to} $d$}
					\State $\ket{q_l,q_{l-d}} \leftarrow \ket{q_{l-d},q_l}\ \cb\ \ket{a_1}$ \Comment{$C\Swap$}
				\EndFor
			\EndIf
			\For{$l \leftarrow \min\Par{d-1,n-1}$ \textbf{to} $0$}
				\State $\ket{a_0} \leftarrow \ket{q_l}$ \Comment{Copy}
				\State $\ket{q_l} \leftarrow \Op{X} \ket{q_l}\ \cb\ \ket{a_0,a_1}$ \Comment{$CC\Op{X}$}
			\EndFor
		\EndFor
		\State $\ket{s} \leftarrow \ket{-s}$ \Comment{Negate (undo neg.)}
		\State $\ket{q} \leftarrow \ket{-q}\ \cb\ \ket{a_1}$ \Comment{$C\Negate$ (undo abs.)}
	\end{algorithmic}
	\caption{Proposed algorithm for in-place bit shifting of $\ket{q}$ by $\ket{s}$.} \label{alg:shift}
\end{algorithm}

\begin{algorithm}[tb]
	\begin{algorithmic}[1]
		\Registers input/output $\ket{q} = \ket{q_{n-1}\ldots q_0}$.
		\For{$k \leftarrow 0$ \textbf{to} $n-1$}
			\State $\ket{q_k} \leftarrow \Op{X} \ket{q_k}$
		\EndFor
		\State $\ket{q} \leftarrow \ket{q+1}$ \Comment{$\Add$ with constant $1$}
	\end{algorithmic}
	\caption{Proposed algorithm for the negation of Two's Complement integral registers. Negation of fixed-point registers is equivalent to negating the register as if it were integral.}
	\label{alg:negate}
\end{algorithm}

\subsection{Bit Shifting}
In order to implement floating-point arithmetic on quantum computers, we require an operator $\Shift\Par{\cdot,n}$ that shifts its input to the right by $n$ places when $n$ is positive, shifts to the left by $-n$ places when $n$ is negative, and leaves the input unchanged when $n$ is $0$. When the input is unsigned, we introduce $0$ values as we shift the register. When the input is signed, the behavior is nearly identical, with the only difference being that a shift to the right fills with values equal to the most significant qubit prior to shifting. For example, if we want to shift a quantum $n$-bit string representing an unsigned number $\ket{a} = \ket{a_{n-1} a_{n-2} \ldots a_0}$ to the right by 1 place, then $\Shift\Par{\cdot,+1}$ must carry out the following operation.
\begin{equation}
	\Shift\Par{\ket{a},+1}\colon \ket{a_{n-1} \ldots a_0} \rightarrow \ket{0\ a_{n-1} \ldots a_1}
\end{equation}
Similarly, a shift to the left by 1 using $\Shift\Par{\cdot,-1}$ should carry out the following operation.
\begin{equation}
	\Shift\Par{\ket{a},-1}\colon \ket{a_{n-1} \ldots a_0} \rightarrow \ket{a_{n-2} \ldots a_0\ 0}
\end{equation}
Finally, when the input is signed and encoded using a Two's Complement representation, then $\Shift\Par{\cdot,+1}$ should carry out the following operation.
\begin{equation}
	\Shift\Par{\ket{a},+1}\colon \ket{a_{n-1} \ldots a_0} \rightarrow \ket{a_{n-1} a_{n-1} \ldots a_1}
\end{equation}

Our approach to implement the $\Shift\Par{\cdot,n}$ operation is detailed in Alg. \ref{alg:shift}. Note that the shifting works when the shift is a superposition of integers $\ket{n}$. To summarize how this approach works, we first note that because a shift is akin to multiplication by $2^n$, if we express $n = \sum_{k=0}^{m-1} n_k 2^k$, then, due to the properties of exponentiation, we can decompose the shift by $n$ into shifts by powers of 2, as can be seen in the following equation.
\begin{equation}
    2^{n} = 2^{\sum_{k=0}^{m-1} n_k 2^k} = \prod_{k = 0}^{m-1} 2^{n_k 2^k}
\end{equation}
Then, for each power of 2, we swap the qubits from most significant to least significant if $n$ contains that power of 2. This is achieved with controlled swaps. We take special care to account for negative inputs when the output is signed, resulting in a second pass on the negated input. The copies to ancilla qubits and negations, controlled or uncontrolled, are merely housekeeping steps to ensure that the correct fill value is used. Note that this is meant to be a proof-of-concept and more efficient ways of performing shifts over superpositions of shift values could be found.

\begin{algorithm}[tb]
\begin{algorithmic}[1]
	\Registers input/output $\ket{q} = \ket{q^e, q^m}$ where exponent $\ket{q^e} = \ket{q^e_{e-1} \ldots q^e_{0}}$ and mantissa $\ket{q^m} = \ket{q^m_{m-1} \ldots q^m_0}$;
	ancillae $\ket{a} = \ket{a_1 a_0}$.
	\State $\ket{a} \leftarrow \ket{0}$ \Comment{Reset}
	\For{$k \leftarrow \Par{m-1}$ \textbf{to} $0$}
		\State $\ket{a_1} \leftarrow \Op{X} \ket{0}\ \cb\ \ket{q^m_k,\overline{a_0}}$ \Comment{$CC\Op{X} \circ \Reset$}
		\State $\ket{a_0} \leftarrow \Op{X} \ket{a_0}\ \cb\ \ket{q^m_k,a_1}$ \Comment{$CC\Op{X}$}
	\EndFor
	\For{$k \leftarrow \Par{e-1}$ \textbf{to} $0$}
		\State $\ket{a_1} \leftarrow \Op{X} \ket{0}\ \cb\ \ket{q^e_k,\overline{a_0}}$ \Comment{$CC\Op{X} \circ \Reset$}
		\State $\ket{q^e_k} \leftarrow \Op{X} \ket{q^e_k}\ \cb\ \ket{a_1}$ \Comment{$CC\Op{X}$}
	\EndFor
\end{algorithmic}
\caption{$\ZeroExp$: Proposed algorithm for setting exponent to $\ket{0}$ when mantissa is zero.} \label{alg:ZeroExp}
\end{algorithm}

\subsection{Setting Exponent to Zero}
Another requirement for implementing floating-point arithmetic is setting the exponent to $\ket{0}$ when the mantissa is equal to $\ket{0}$. This procedure requires two ancilla qubits $\ket{a_1 a_0}$. We use ancilla $\ket{a_0}$ to keep track of whether the mantissa is non-zero and ancilla $\ket{a_1}$ for temporary calculations. We determine whether the mantissa is non-zero by scanning the mantissa qubits, whereby at each step, we reset $\ket{a_1}$, invert $\ket{a_1}$ if $\ket{a_0}$ is zero and the mantissa qubit is non-zero, and invert $\ket{a_0}$ if $\ket{a_1}$ is active and the mantissa qubit is also active. We then scan the exponent qubits, whereby at each step, we reset $\ket{a_1}$, invert $\ket{a_1}$ if the exponent qubit is active and $\ket{a_0}$ is inactive, and then we invert the exponent qubit if $\ket{a_1}$ is active. Algorithm \ref{alg:ZeroExp} describes this procedure in more detail.

\subsection{Floating-Point Encoding}
We choose to represent superpositions of floating point values by separating each floating point number into a Two's Complement signed integer exponent and a Two's Complement $\Par{f+1,f}$ signed fixed-point mantissa, denoted with superscripts $e$ and $m$, respectively. For example, given a prescribed exponent width of 5 and a mantissa width of 11 for register $\ket{a}$,
\begin{equation}
	\ket{a} = \ket{a^e a^m} = \ket{a_{15} a_{14} \ldots a_{11} a_{10} \ldots a_{0}}\text{,}
\end{equation}
where exponent $\ket{a^e} = \ket{a_{15}\ldots a_{11}}$ and mantissa $\ket{a^m} = \ket{a_{10}\ldots a_{0}}$. Note that this encoding does not exactly mimic IEEE-754 and was chosen because it builds on our previous work on Two's Complement signed fixed-point encodings \cite{serralles2024quantum}. Fig. \ref{fig:floating} illustrates the difference between the IEEE-754 convention and our similar encoding. Extending our work to One's Complement numbers is a trivial exercise and would result in operations with similar complexity, meaning that our approach could be used to exactly mimic IEEE-754.

\begin{algorithm}[tb!]
	\begin{algorithmic}[1]
		\Registers inputs $\ket{q} = \ket{q^e, q^m} = \ket{q^e_{e-1} \ldots q^e_0, q^m_{m-1} \ldots q^m_0}$ and $\ket{r} = \ket{r^e, r^m}$; output $\ket{s^e, s^m}$; and ancillae $\ket{a} = \ket{a_{n-1} a_{n-2} \ldots a_0}$, where $n = \max\Par{m,7}$.
		\Let $\ket{s^{m,m-1}} \coloneq \ket{s^m a_{n-1} \ldots a_{n-m+1}}$, $\ket{s^{m,1}} \coloneq \ket{s^m a_{n-1}}$, and $\ket{a^s} \coloneq \ket{a_{n-4} a_{n-5} a_{n-6} a_{n-7}}$.
		\State $\ket{s^{m,m-1}} \leftarrow \ket{s^{m,m-1} + q^m \cdot r^m}$ \Comment{$\FMA$}
		\State $\ket{a_{n-2}} \leftarrow \ket{s^m_0}$ \Comment{$\Copy$}
		\State $\ket{s^{m,1}} \leftarrow \ket{-s^{m,1}}\ \cb\ \ket{a_{n-2}}$ \Comment{$C\Negate$ (abs. value)}
		\State $\ket{a_{n-3}} \leftarrow \ket{a_{n-2}}$ \Comment{$\Copy$}
		\State $\ket{a_{n-3}} \leftarrow \Op{X} \ket{a_{n-3}}$
		\State $\ket{s^{m,1}} \leftarrow \Shift\Par{\ket{s^{m,1}},\ket{a_{n-3}},\ket{a^s}}$ \Comment{$\Shift$ if $\left\vert q^m \right\vert < 0.5$}
		\State $\ket{s^e} \leftarrow \ket{s^e+a_{n-3}}$ \Comment{Adjust $q^e$ with $\Add$}
		\State $\ket{s^{m,1}} \leftarrow \ket{-s^{m,1}}\ \cb\ \ket{a_{n-2}}$ \Comment{$C\Negate$ (undo abs. value)}
		\State $\ket{a_{n-1} a_{n-2}} \leftarrow \ket{0}$ \Comment{$\Reset$}
		\Let $\ket{s^{1,e}} \coloneq \ket{a_{n-1} s^e}$.
		\State $\ket{a_{n-1}} \leftarrow \Op{X} \ket{a_{n-1}}\ \cb\ \ket{s^e_0}$ \Comment{$C\Op{X}$ (extend sign)}
		\State $\ket{s^{1,e}} \leftarrow \ket{s^{1,e} + q^e}$ \Comment{$\Add$}
		\State $\ket{s^{1,e}} \leftarrow \ket{s^{1,e} + r^e}$ \Comment{$\Add$}
		\State $\ket{a_{n-2}} \leftarrow \Op{X} \ket{a_{n-2}}\ \cb\ \ket{a_{n-1}, \overline{s^e_0}}$ \Comment{$CC\Op{X}$}
		\For{$k \leftarrow \Par{m-1}$ \textbf{to} $0$} \Comment{Underflow correction}
			\State $\ket{a_{n-1}} \leftarrow \ket{s^m_k}$ \Comment{$\Copy$}
			\State $\ket{s^m_k} \leftarrow \Op{X} \ket{s^m_k}\ \cb\ \ket{a_{n-1},a_{n-2}}$ \Comment{$CC\Op{X}$}
		\EndFor
		\State $\ket{s} \leftarrow \ZeroExp\Par{\ket{s},\ket{a^s}}$ \Comment{Zero exponent if mantissa is 0}
	\end{algorithmic}
	\caption{Proposed algorithm for multiplication of floating-point registers.}
	\label{alg:mult}
\end{algorithm}

\subsection{Floating-Point Multiplication} \label{sec:multiplication} % TODO: Add extra logic relating to zeroing out the exponent if mantissa equals 0, and logic to prevent underflow/overflow
For convenience, multiplication is not done in-place as for the case of fixed-point numbers \cite{serralles2024quantum} due to the otherwise required $\Shift$ operations. We denote the multiplication operation as $\Mult\Par{\cdot}$. We denote the mantissa and exponent of a number using superscripts $m$ and $e$, respectively. Given two input numbers $a = a^m \cdot 2^{a^e}$ and $b = b^m \cdot 2^{b^e}$, the output number $c = c^m \cdot 2^{c^e}$ set to the product of $a$ and $b$ is given by
\begin{equation}
	c^m \cdot 2^{c^e} = \Mult\Par{a,b} = a^m \cdot b^m \cdot 2^{a^e+b^e}\text{.}
\end{equation}
Therefore, $c^m = a^m \cdot b^m$ and $c^e = a^e+b^e$. We use the fixed-point $\FMA$ operation to compute the mantissa and the $\Add$ operation to compute the exponent. We prepend an ancilla qubit to $c^m$ to allow for overflow in the mantissa. If overflow occurs, then we shift the mantissa by 1 to the right using the $\Shift$ operation, and then add 1 to the exponent $c^e$, which is done by controlling using a copy of the overflow qubit in another ancilla qubit. We also prepend an ancilla qubit to the exponent to allow for underflow, in which case the output is set to $0$. Finally, we set the exponent to $0$ if the mantissa is $0$ using the $\ZeroExp$ gate (Alg. \ref{alg:ZeroExp}). We present a full algorithmic description of the multiplication operation in Algorithm \ref{alg:mult}.

\begin{algorithm}[tb!]
	\begin{algorithmic}[1]
		\Registers inputs $\ket{q} = \ket{q^e,\,q^m} = \ket{q^e_{e-1} \ldots q^e_0,\,q^m_{m-1}\ldots q^m_0}$ and $\ket{r} = \ket{r^e, r^m}$; output $\ket{s^e, s^m}$; and ancillae $\ket{a} = \ket{a_7 a_6 \ldots a_0}$.
		\Let $\ket{s^{m,1}} \coloneq \ket{s^m a_2}$, $\ket{s^{m,2}} \coloneq \ket{a_1 s^{m,1}}$, and $\ket{a^\prime} \coloneq \ket{a_6 a_5 a_4 a_3}$.
		\State $\ket{a} \leftarrow \ket{0}$ \Comment{$\Reset$}
		\State $\ket{s^e} \leftarrow \ket{q^e}$ \Comment{$\Copy$}
		\State $\ket{s^e} \leftarrow \ket{s^e - r^e}$ \Comment{$\Add^\dagger$}
		\State $\ket{a_0} \leftarrow \Op{X}\ket{a_0}\ \cb\ \ket{s^e_0}$ \Comment{$C\Op{X}$}
		\State $\ket{s^m} \leftarrow \ket{0}$ \Comment{$\Reset$}
		\For{$k \leftarrow \Par{m-1}$ \textbf{to} $0$}
			\State $\ket{s^m_k} \leftarrow \Op{X} \ket{s^m_k}\ \cb\ \ket{a_0, q^m_k}$ \Comment{$CC\Op{X}$}
		\EndFor
		\State $\ket{a_0} \leftarrow \Op{X} \ket{a_0}$
		\For{$k \leftarrow \Par{m-1}$ \textbf{to} $0$}
			\State $\ket{s^m_k} \leftarrow \Op{X} \ket{s^m_k}\ \cb\ \ket{a_0, r^m_k}$ \Comment{$CC\Op{X}$}
		\EndFor
		\State $\ket{s^e} \leftarrow \ket{-s^e}\ \cb\ \ket{\overline{a_0}}$ \Comment{$C\Negate$}
		\State $\ket{s^{m,1}} \leftarrow \Shift\Par{\ket{s^{m,1}}, \ket{s^e}, \ket{a^\prime}}$ \Comment{Alg. \ref{alg:shift}}
		\State $\ket{a_1} \leftarrow \Op{X} \ket{a_1}\ \cb\ \ket{s^m_0}$ \Comment{$C\Op{X}$}
		\State $\ket{s^{m,2}} \leftarrow \ket{s^{m,2} + q^m}\ \cb\ \ket{a_0}$ \Comment{$C\Add$}
		\State $\ket{s^{m,2}} \leftarrow \ket{s^{m,2} + r^m}\ \cb\ \ket{\overline{a_0}}$ \Comment{$C\Add$}
		\State $\ket{a_7} \leftarrow \ket{a_1}$ \Comment{$\Copy$}
		\State $\ket{s^{m,2}} \leftarrow \ket{-s^{m,2}}\ \cb\ \ket{a_7}$ \Comment{$\Negate$}
		\State $\ket{s^e} \leftarrow \ket{1}$ \Comment{$\Copy$}
		\State $\ket{a^\prime} \leftarrow \ket{0}$ \Comment{$\Reset$}
		\For{$k \leftarrow m$ \textbf{to} $0$}
			\State $\ket{s^e} \leftarrow \ket{s^e+1}\ \cb\ \ket{\overline{s^{m,1}_k}, \overline{a_7}}$
			\State $\ket{a_6} \leftarrow \ket{a_7}$ \Comment{$\Copy$}
			\State $\ket{a_7} \leftarrow \Op{X} \ket{a_7}\ \cb\ \ket{s^{m,1}_k, \overline{a_6}}$
		\EndFor
		\State $\ket{s^{m,2}} \leftarrow \Shift\Par{\ket{s^{m,2}},s^e,a^\prime}$ \Comment{Alg. \ref{alg:shift}}
		\State $\ket{s^m} \leftarrow \ket{-s^m}\ \cb\ \ket{a_7}$ \Comment{$C\Negate$}
		\State $\ket{s^e} \leftarrow \ket{s^e + r^e}\ \cb\ \ket{a_0}$ \Comment{$C\Add$}
		\State $\ket{s^e} \leftarrow \ket{s^e + q^e}\ \cb\ \ket{\overline{a_0}}$ \Comment{$C\Add$}
		\State $\ket{s} \leftarrow \ZeroExp\Par{\ket{s},\ket{a}}$ \Comment{Alg. \ref{alg:ZeroExp}}
	\end{algorithmic}
	\caption{Proposed algorithm for addition of floating-point registers.}
	\label{alg:add}
\end{algorithm}

\subsection{Floating-Point Addition} \label{sec:addition}
Floating-point addition is considerably more difficult to implement than multiplication because of possible cancellations in the mantissas, and because the inputs must be shifted prior to addition to ensure that the addition is performed under the same exponent value. Let us assume that we want to add two floating point registers $\ket{a} = \ket{a^e a^m}$ and $\ket{b} = \ket{b^e b^m}$ and to store the output in register $\ket{c} = \ket{c^e c^m}$. We start by setting the output register $\ket{c}$ to the ground state using the reset approach presented in this work and copy $\ket{a^e}$ to $\ket{c^e}$ using controlled $\Op{X}$ gates. We then subtract the exponents $\ket{c^e}$ and $\ket{b^e}$ using $\Add^\dagger\Par{\ket{c^e},\ket{b^e}}$. We then use this value to copy the appropriate input mantissa to the output mantissa using doubly controlled $\Op{X}$ gates. If the difference is negative, then we copy the second input; otherwise, we copy the first input. We then set the output exponent to its absolute value by using controlled negation, and shift the output mantissa to the right by using $\Shift\Par{\ket{c^m},\ket{c^e}}$. After shifting, we add the two mantissas with the operation $\Add\Par{\ket{c^m},\ket{b^m}}$. Then, we scan the mantissa from most to least significant qubit, subtracting 1 from an ancilla integer counter for every zero qubit until the first non-zero qubit is encountered. Because this operation must work for all possible combinations of inputs, we implement the scanning procedure by controlling with an ancilla qubit that indicates whether a non-zero qubit has been encountered. Using this ancilla counter, we shift the output mantissa as appropriate and then add this counter to the exponent of the output, which yields the correct value in the output register.

\subsection{Higher-Order Floating-Point Operations}
We can combine addition and multiplication to implement other operations. For example, we can use the same Newton's Method approach that we used for fixed-point division in \cite{serralles2024quantum} to carry out floating-point division. We can also implement exponentiation for small arguments by expressing the exponential function in terms of its Taylor series, and then factorizing the polynomial approximation using Horner's Method for Polynomial Evaluation, as summarized in the following equation for polynomial approximation order $N$.
\begin{equation}
\begin{aligned}
	\exp\Par{x} &\approx \sum_{k=0}^{N} \frac{x^k}{k!} \\
				&\approx 1+x\cdot\Par{1+\frac{x}{2}\cdot\Par{1+\ldots \Par{1+\frac{x}{N}}}}
\end{aligned}
\end{equation}
For sufficiently high $N$ (on the order of 10-12), this approximation is accurate to the precision of the floating-point representation. This technique could be extended to larger arguments by storing a look-up table of values (classically) for each power of 2 in a range. We can also extend this approach to the trigonometric functions $\sin\Par{x}$ and $\cos\Par{x}$, and to variants $\sin\Par{\pi x}$ and $\cos\Par{\pi x}$. Once we implement these functions, we can derive the inverse trigonometric functions $\arcsin\Par{x}$ and $\arccos\Par{x}$, other trigonometric functions like $\tan\Par{x}$, the natural logarithm $\log\Par{x}$, and so on.

\begin{table*}[tbp!]
	\centering
	\includegraphics[]{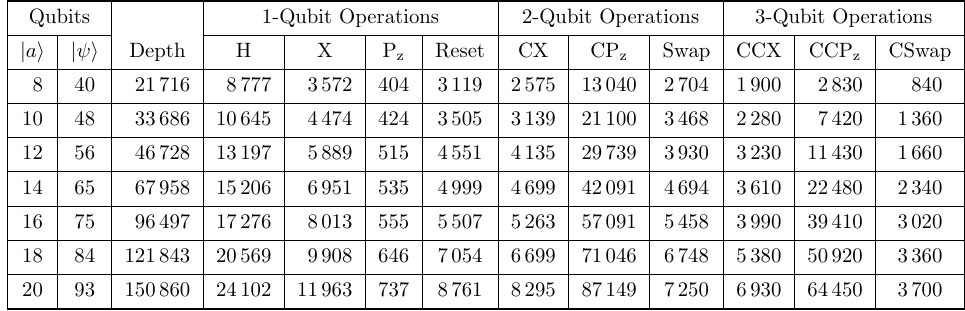}
	\caption{Summary of $\Recip$ resource utilization for every register width. The first, second, and third columns list the number of qubits per register, the total number of qubits in the wavefunction, and the circuit depth, respectively. The remaining columns list the number of operations, in order of increasing gate width.\label{tab:div}}
\end{table*}

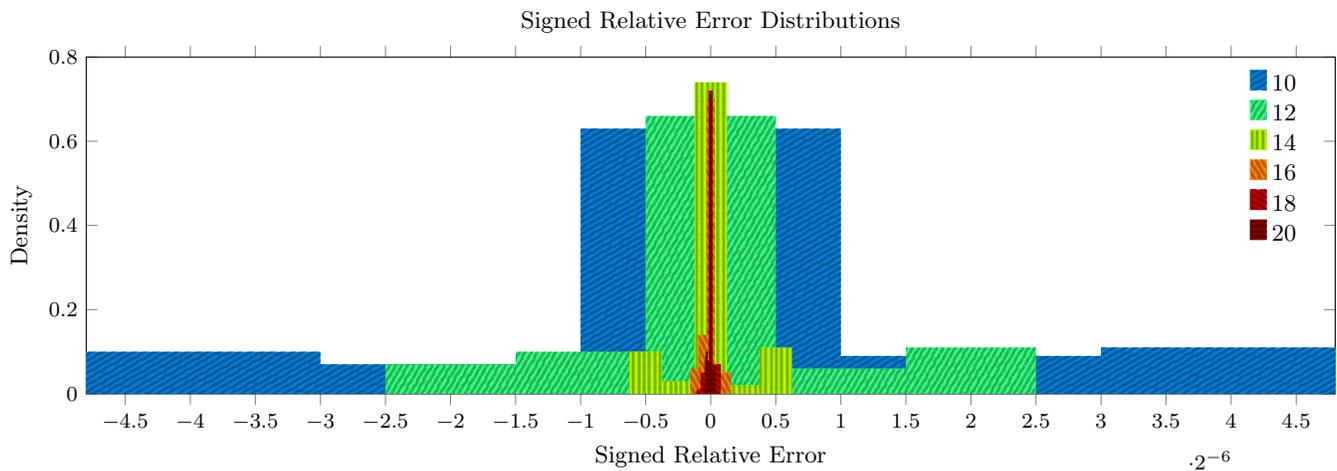
\begin{figure*}[ht!]
    \centering
	% Recommended preamble:
\captionsetup[sub]{font=small}
\begin{tikzpicture}
\begin{axis}[small, ybar={0pt}, bar shift={0pt}, title={Signed Relative Error Distributions}, xlabel={Signed Relative Error}, ylabel={Density}, ymin={0}, ymax={0.8}, title style={font={\small}}, scaled x ticks={real:0.015625}, xtick scale label code/.code={{$\cdot 2^{-6}$}}, xticklabel style={/pgf/number format/fixed, /pgf/number format/precision={1}}, xtick distance={0.0078125}, width={7.16in}, height={2.3866666666666667in}, single ybar legend, legend style={draw={none}, fill={none}, inner ysep={2pt}, cells={anchor={west}}}, colormap/bluered, cycle multiindex* list={{[colors of colormap={100,300,...,1000}] \nextlist pattern={Lines[yshift=.5pt,angle=30,distance=2pt,line width=1pt]},pattern={Lines[yshift=.5pt,angle=60,distance=2pt,line width=1pt]},pattern={Lines[yshift=.5pt,angle=90,distance=2pt,line width=1pt]},pattern={Lines[yshift=.5pt,angle=120,distance=2pt,line width=1pt]},pattern={Lines[yshift=.5pt,angle=150,distance=2pt,line width=1pt]},pattern={Lines[yshift=.5pt,angle=180,distance=2pt,line width=1pt]},pattern={Lines[yshift=.5pt,angle=210,distance=2pt,line width=1pt]},pattern={Lines[yshift=.5pt,angle=240,distance=2pt,line width=1pt]},pattern={Lines[yshift=.5pt,angle=270,distance=2pt,line width=1pt]},pattern={Lines[yshift=.5pt,angle=300,distance=2pt,line width=1pt]},pattern={Lines[yshift=.5pt,angle=330,distance=2pt,line width=1pt]},pattern={Lines[yshift=.5pt,angle=360,distance=2pt,line width=1pt]},pattern=grid,pattern=crosshatch,pattern=dots \nextlist preaction={fill=.} \nextlist pattern color=.!70!black}}]
    \addplot+[bar width={0.03125}, draw={none}]
        table[row sep={\\}]
        {
            \\
            -0.0625  0.1  \\
            -0.03125  0.07  \\
            0.0  0.63  \\
            0.03125  0.09  \\
            0.0625  0.11  \\
        }
        ;
    \addlegendentry {10}
    \addplot+[bar width={0.015625}, draw={none}]
        table[row sep={\\}]
        {
            \\
            -0.03125  0.07  \\
            -0.015625  0.1  \\
            0.0  0.66  \\
            0.015625  0.06  \\
            0.03125  0.11  \\
        }
        ;
    \addlegendentry {12}
    \addplot+[bar width={0.00390625}, draw={none}]
        table[row sep={\\}]
        {
            \\
            -0.0078125  0.1  \\
            -0.00390625  0.03  \\
            0.0  0.74  \\
            0.00390625  0.02  \\
            0.0078125  0.11  \\
        }
        ;
    \addlegendentry {14}
    \addplot+[bar width={0.0009765625}, draw={none}]
        table[row sep={\\}]
        {
            \\
            -0.001953125  0.06  \\
            -0.0009765625  0.14  \\
            0.0  0.7  \\
            0.0009765625  0.05  \\
            0.001953125  0.05  \\
        }
        ;
    \addlegendentry {16}
    \addplot+[bar width={0.00048828125}, draw={none}]
        table[row sep={\\}]
        {
            \\
            -0.00146484375  0.01  \\
            -0.0009765625  0.05  \\
            -0.00048828125  0.08  \\
            0.0  0.72  \\
            0.00048828125  0.07  \\
            0.0009765625  0.07  \\
            0.00146484375  0.0  \\
        }
        ;
    \addlegendentry {18}
    \addplot+[bar width={0.000244140625}, draw={none}]
        table[row sep={\\}]
        {
            \\
            -0.000732421875  0.01  \\
            -0.00048828125  0.1  \\
            -0.000244140625  0.07  \\
            0.0  0.71  \\
            0.000244140625  0.05  \\
            0.00048828125  0.06  \\
            0.000732421875  0.0  \\
        }
        ;
    \addlegendentry {20}
\end{axis}
\end{tikzpicture}
	\caption{Distribution of signed error of output of reciprocal gate when compared with expected output. We assessed performance using register widths of 10, 12, 14, 16, 18, and 20 qubits, with corresponding mantissa widths of 6, 7, 9, 11, 12, and 13 and exponent widths 4, 5, 5, 5, 6, and 7. Each reciprocal gate was tested with 100 samples from a normal distribution $\mathcal{N}\Par{0,5}$ and for $10$ Newton iterations. We discarded samples that did not have a representable reciprocal, such as $0$.}
    \label{fig:div:stats}
\end{figure*}

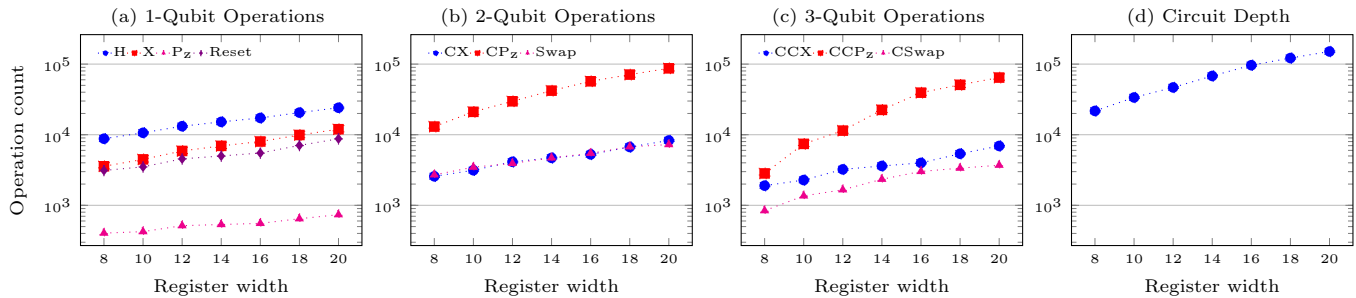
\begin{figure*}[t]
    \centering
	% Recommended preamble:
\captionsetup[sub]{font=scriptsize}
\begin{tikzpicture}
\begin{groupplot}[group style={group size={4 by 1}, ylabels at={edge left}, horizontal sep={0.65cm}, vertical sep={0.5cm}}, legend image code/.code={{ \draw[mark repeat=2,mark phase=2] plot coordinates { (0cm,0cm) (0.2cm,0cm) (0.4cm,0cm) }; }}, footnotesize, title style={font={\scriptsize}, text width={1.4730708661417324in}, yshift={-1mm}}, label style={font={\scriptsize}}, tick label style={font={\tiny}}, legend pos={north west}, legend image post style={scale={0.5}}, legend columns={-1}, legend style={draw={none}, fill={none}, inner ysep={2pt}, cells={anchor={west}}, nodes={inner sep={0pt}, text depth={0.15em}}, font={\tiny}}, scale only axis, width={1.4730708661417324in}, height={1.1048031496062993in}, ymajorgrids={true}, xlabel={Register width}, ylabel={Operation count}, cycle multiindex* list={mark=*,mark=square*,mark=triangle*,mark=diamond*\nextlist blue,red,magenta,violet\nextlist dotted}, ymin={269.3333333333333}, ymax={261447}]
    \nextgroupplot[xmode=normal,ymode=log, title={\subcaption{1-Qubit Operations \label{fig:div:counts:w1}}}]
    \addplot+
        table[row sep={\\}]
        {
            \\
            8  8777  \\
            10  10645  \\
            12  13197  \\
            14  15206  \\
            16  17276  \\
            18  20569  \\
            20  24102  \\
        }
        ;
    \addlegendentry {H}
    \addplot+
        table[row sep={\\}]
        {
            \\
            8  3572  \\
            10  4474  \\
            12  5889  \\
            14  6951  \\
            16  8013  \\
            18  9908  \\
            20  11963  \\
        }
        ;
    \addlegendentry {X}
    \addplot+
        table[row sep={\\}]
        {
            \\
            8  404  \\
            10  424  \\
            12  515  \\
            14  535  \\
            16  555  \\
            18  646  \\
            20  737  \\
        }
        ;
    \addlegendentry {P\textsubscript{z}}
    \addplot+
        table[row sep={\\}]
        {
            \\
            8  3119  \\
            10  3505  \\
            12  4551  \\
            14  4999  \\
            16  5507  \\
            18  7054  \\
            20  8761  \\
        }
        ;
    \addlegendentry {Reset}
    \nextgroupplot[xmode=normal,ymode=log, title={\subcaption{2-Qubit Operations \label{fig:div:counts:w2}}}]
    \addplot+
        table[row sep={\\}]
        {
            \\
            8  2575  \\
            10  3139  \\
            12  4135  \\
            14  4699  \\
            16  5263  \\
            18  6699  \\
            20  8295  \\
        }
        ;
    \addlegendentry {CX}
    \addplot+
        table[row sep={\\}]
        {
            \\
            8  13040  \\
            10  21100  \\
            12  29739  \\
            14  42091  \\
            16  57091  \\
            18  71046  \\
            20  87149  \\
        }
        ;
    \addlegendentry {CP\textsubscript{z}}
    \addplot+
        table[row sep={\\}]
        {
            \\
            8  2704  \\
            10  3468  \\
            12  3930  \\
            14  4694  \\
            16  5458  \\
            18  6748  \\
            20  7250  \\
        }
        ;
    \addlegendentry {Swap}
    \nextgroupplot[xmode=normal,ymode=log, title={\subcaption{3-Qubit Operations \label{fig:div:counts:w3}}}]
    \addplot+
        table[row sep={\\}]
        {
            \\
            8  1900  \\
            10  2280  \\
            12  3230  \\
            14  3610  \\
            16  3990  \\
            18  5380  \\
            20  6930  \\
        }
        ;
    \addlegendentry {CCX}
    \addplot+
        table[row sep={\\}]
        {
            \\
            8  2830  \\
            10  7420  \\
            12  11430  \\
            14  22480  \\
            16  39410  \\
            18  50920  \\
            20  64450  \\
        }
        ;
    \addlegendentry {CCP\textsubscript{z}}
    \addplot+
        table[row sep={\\}]
        {
            \\
            8  840  \\
            10  1360  \\
            12  1660  \\
            14  2340  \\
            16  3020  \\
            18  3360  \\
            20  3700  \\
        }
        ;
    \addlegendentry {CSwap}
    \nextgroupplot[xmode=normal,ymode=log, title={\subcaption{Circuit Depth \label{fig:div:counts:depth}}}]
    \addplot+
        table[row sep={\\}]
        {
            \\
            8  21716  \\
            10  33686  \\
            12  46728  \\
            14  67958  \\
            16  96497  \\
            18  121843  \\
            20  150860  \\
        }
        ;
\end{groupplot}
\end{tikzpicture}
	\caption{Number of operations per Recip operation as a function of floating-point register width. Figs.~(\protect\subref{fig:div:counts:w1})-(\protect\subref{fig:div:counts:w3}) shows counts of 1-, 2-, and 3-qubit operations, respectively, as a function of register width, while Fig.~(\protect\subref{fig:div:counts:depth}) shows the estimate for the overall circuit depth growth as a function of register width.}
    \label{fig:div:counts}
\end{figure*}

\begin{table*}[tbp!]
	\centering
	\includegraphics[width=\textwidth]{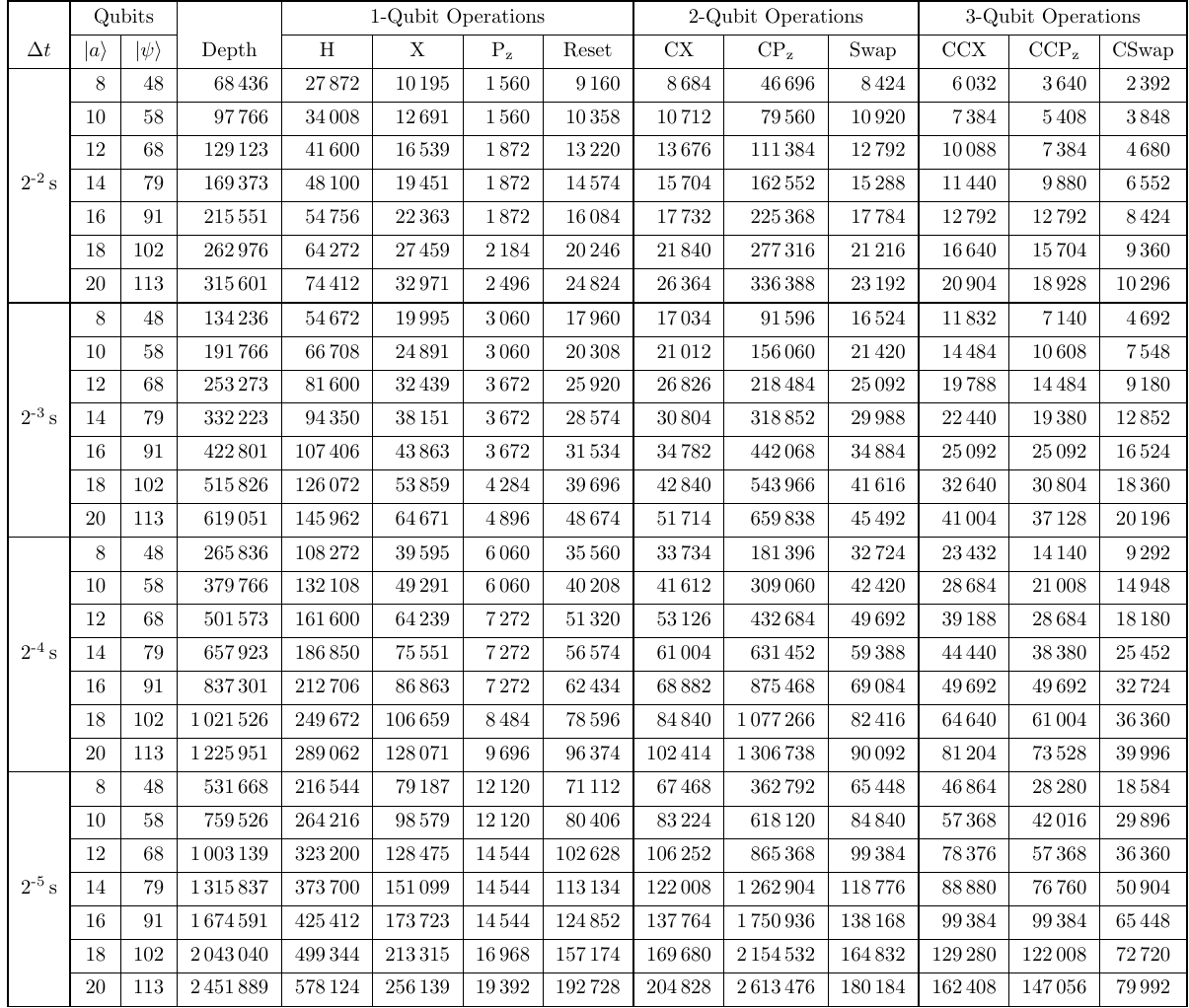}
	\caption{Summary of resource utilization for every ODE simulation. The first, second, and third columns list the time integration step size, the number of qubits per register, and the total number of qubits in the wavefunction, respectively. The remaining columns list the number of operations, in order of increasing gate width.\label{tab:ode}}
\end{table*}

\begin{figure*}[t]
	\centering
	\input{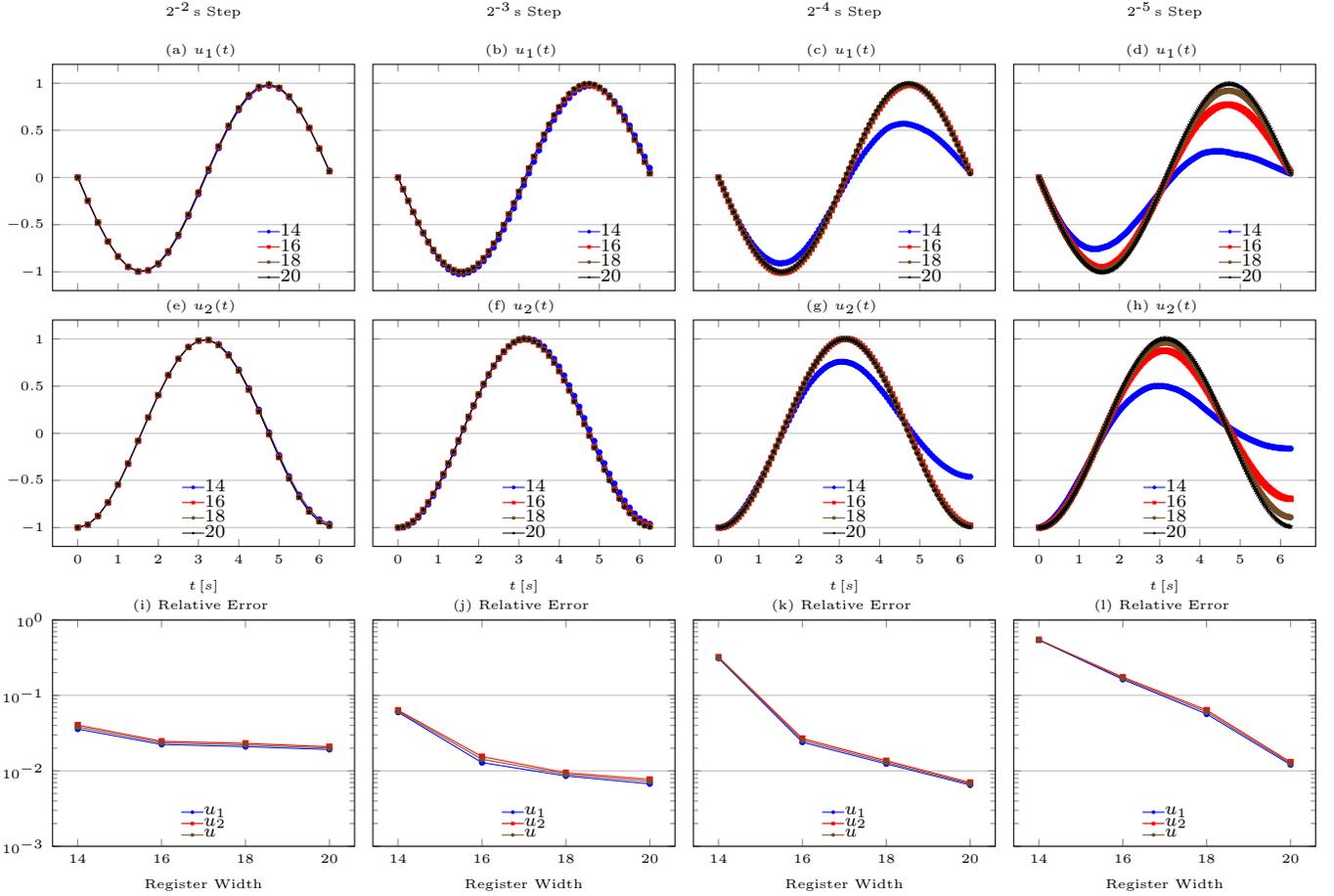}
    \caption{Results from the solution of the tested system of ordinary differential equations for 14-, 16-, 18-, and 20-qubit registers. Each column in the figure corresponds to a different timestep $\Delta t$ used when integrating the system. The top and middle rows show the evolution of $u_1$ and $u_2$, respectively, for different $\Delta t$. Each data point is represented using a circular marker, resulting in a thicker appearance as $\Delta t$ is refined. The bottom row shows the relative error with respect to the analytical solution as a function of register widths, for different $\Delta t$.}
	\label{fig:ode:evol}
\end{figure*}

\begin{figure*}[t]
	\centering
	\input{plots/ode_counts.tikz}
	\caption{Number of operations per simulation of the ordinary differential equation system of Eq.~\eqref{eq:ode:system}. We simulated the system for one period, while varying the step from $\unit[2^{-2}]{s}$ to $\unit[2^{-5}]{s}$. Figs.~(\protect\subref{fig:ode:counts:w1sp2})-(\protect\subref{fig:ode:counts:w1sp5}) show count of 1-qubit operations for each step size, as a function of register width. Similarly, Figs.~(\protect\subref{fig:ode:counts:w2sp2})-(\protect\subref{fig:ode:counts:w2sp5}) and Figs.~(\protect\subref{fig:ode:counts:w3sp2})-(\protect\subref{fig:ode:counts:w3sp5}) show counts of 2- and 3-qubit operations, respectively, for each step size and as a function of register width. Figs.~(\protect\subref{fig:ode:counts:depth:sp2})-(\protect\subref{fig:ode:counts:depth:sp5}) show the overall circuit depth for each step size and as a function of register width.}
\label{fig:ode:counts} 
\end{figure*}

\section{Methods}
As a proof-of-concept, we used our floating-point framework to replicate the analysis of the fixed-point reciprocal operation that was presented in \cite{serralles2024quantum}, all in simulation on a classical computer (Apple M2 SoC with 8 cores and \unit[24]{GiB} RAM, 2022). The reciprocal operation approximates the reciprocal using Newton's Method for root finding. The main difference from the fixed-point case is in how we obtain the initial guess: Given an input, we simply set the mantissa to $\pm 1$ and negate the input's exponent. We carried out the same analysis for 10-, 12-, 14-, 16- and 18-qubit floating point registers.  

Using our proposed floating-point encoding scheme, we also replicated the numerical solution of the system of ordinary differential equations that we had previously simulated using fixed-point registers in \cite{serralles2024quantum}:
\begin{equation} \label{eq:ode:system}
	\deriv{\vec{u}}{t} = \deriv{}{t}\!\begin{bmatrix} u_1\Par{t} \\ u_2\Par{t} \end{bmatrix} = \begin{bmatrix} \hphantom{-}0 & 1 \\ -1 & 0 \end{bmatrix} \begin{bmatrix} u_1\Par{t} \\ u_2\Par{t} \end{bmatrix}
\end{equation}
with initial condition $\vec{u}\Par{0^+}^\T = \vec{u}_0^\T = \begin{bmatrix} 0 & -1 \end{bmatrix}$. The analytical solution of this system is given by
\begin{equation} \label{eq:ode:analytical}
	\vec{u}\Par{t} = -\begin{bmatrix} \sin\Par{t} \\ \cos\Par{t} \end{bmatrix}\Theta\Par{t}\text{,}
\end{equation}
where $\Theta\Par{t}$ is the Heaviside step function. We solve the system using 14-, 16-, 18-, and 20-qubit floating-point registers. For each register width, we employed time steps of $\unit[2^{-2}]{s}$, $\unit[2^{-3}]{s}$, $\unit[2^{-4}]{s}$, and $\unit[2^{-5}]{s}$ for the evolution of the system using the trapezoidal rule for numerical integration. For each time step, we simulated the system for approximately $\unit[2\pi]{s}$, corresponding to the period of the analytical solution in Eq. \eqref{eq:ode:analytical}. For time step $\Delta t$, we approximated the analytic solution of the system with the following explicit trapezoidal rule update step.
\begin{equation} \label{eq:ode:trapzstep}
    \mathbf{u}_{k+1} \leftarrow \frac{1}{1+\frac{\Delta t^2}{4}} \begin{bmatrix}
        1-\frac{\Delta t^2}{4} & \Delta t \\
        -\Delta t & 1-\frac{\Delta t^2}{4}
	\end{bmatrix} \mathbf{u}_k\text{,}
\end{equation}

In order to assess the performance of our proposed methods, we counted the total number of Clifford and rotation operations when simulating these circuits. Note that we performed this without decomposing Toffoli gates as is done to evaluate Clifford+T gate-based approaches. In those cases, authors often use T gate count and depth as metrics for the complexity of quantum circuits, in addition to Toffoli depth, Toffoli count, and qubit count. For QFT-based approaches like ours, common performance metrics instead include QFT gate count and non-Clifford (rotation) gate count, excluding those used in the QFT routine. In some instances, the evaluation metrics used for QFT-based designs can be related to those of Clifford+T-based designs \cite{paler2022quantum}. Here, we chose to report the raw count of elementary gates and their controlled variants because counting only QFT operations can obscure the true complexity of the circuit, since an $n$-qubit QFT circuit requires $n(n-1) \div 2$ non-Clifford controlled phase gates.

\section{Results}
Fig.~\ref{fig:div:stats} shows the signed relative error distributions of the output of the $\Recip$ operation for different register widths and for 100 samples from a Gaussian distribution with mean of 0 and standard deviation of 5. We discarded samples from the Gaussian distribution whose reciprocals are not representable, such as $\nicefrac{1}{0}$. We observed an exponential decay in the error as we increased the register widths, which matched our expectations. Fig.~\ref{fig:div:counts} illustrates the gate resources required for implementing the reciprocal operation across various qubit register sizes. The total number of 1-, 2-, and 3-qubit gates increased approximately linearly with respect to the width of the register. The proportion by operation type also varies: there are relatively fewer Reset gates in the 1-qubit category, fewer CNOT gates in the 2-qubit category, and fewer CSWAP gates in the 3-qubit category. Among these, the number of Toffoli (CCX) gates offers a more realistic and informative measure of circuit complexity, as further elaborated in the discussion.

Fig.~\ref{fig:ode:evol} shows the results of the numerical experiment, with each column corresponding to a different time step. The top, middle, and bottom rows show the evolution of component $u_1$, the evolution of component $u_2$, and the relative error in the $\ell_2$ sense, respectively, for the prescribed register widths. The relative error in each case exhibits two regimes: in the first regime, limited precision in the mantissa dominates, whereas in the second regime, the trapezoidal rule approximation error dominates. Fig.~\ref{fig:ode:counts} shows the total number of operations used in each ODE simulation. We observed a roughly linear relationship between the register widths and the number of operations, except for rotations P\textsubscript{z}, which increased quadratically due to our use of QFT-based fixed-point arithmetic. As the time step is halved, we observe a corresponding doubling in the number of operations, as expected.

\section{Discussion}
% TODO: write discussion section, which should
% * summarize main aim of the work
% * discuss the results relative to the results in previous work
% * list the limitations of this work and comment about possible ways to overcome them in future work
While fixed-point arithmetic is often simpler to implement on quantum hardware than floating-point arithmetic, the latter provides several key advantages, including a wider dynamic range, improved control over relative error, and natural compatibility with QFT-based circuit components. These benefits make floating-point encoding particularly appealing for quantum algorithms that must operate over a broad range of magnitudes and require high precision. Additionally, floating-point designs often result in shallower circuits with fewer qubits, which is an important consideration for both near-term quantum devices and future fault-tolerant systems.

In this work, we developed a floating-point arithmetic framework that minimizes the number of required ancilla qubits. We accomplished this by leveraging our previously proposed arithmetic primitives \cite{serralles2024quantum} and introducing an ancilla reuse strategy to optimize quantum resource efficiency. In the context of division or reciprocation, our approach used 13 ancilla qubits for 20-qubit reciprocation. A 32-qubit single-precision example would require 23 ancilla qubits, a dramatic reduction from the thousands of qubits that were required with approaches proposed in previous work \cite{gayathri2021a,gayathri2021b}.

Implementing floating-point multiplication was relatively straightforward, as it directly reused fixed-point addition and multiplication subroutines that we had previously developed. Floating-point addition, however, required additional circuit complexity. In fact, the mantissas must be aligned by shifting to a common exponent before addition, which introduced extra gate depth and an increased number of ancilla qubits. Our proposed floating-point encoding also produced significant improvements in numerical accuracy when compared with our fixed-point encoding. For example, we observed a mean error reduction from $2^{-4}$ to $2^{-10}$ in our reciprocal computation benchmarks, compared to our previous results for fixed-point arithmetic \cite{serralles2024quantum}. When solving the differential equation using fixed-point arithmetic \cite{serralles2024quantum}, for time step $\Delta t = 2^{-4}$, the smallest relative error achieved was approximately $2^{-6}$ using fractional bits $f=12$, corresponding to a register width of $2f+1=25$ qubits. In contrast, as shown in Fig.~\ref{fig:ode:evol:error:sp4}, the floating-point representation achieved a lower relative error of $2^{-8}$ using only $\sim$ 20 qubits per register, demonstrating better precision with fewer resources.

Our approach has limitations that prevent its use on present-day quantum hardware. In particular, our approach depends on the use of doubly controlled single-qubit gates and singly controlled Swap gates, which would be difficult to implement efficiently on current quantum hardware. We anticipate, however, that as quantum architectures continue to evolve, these three-qubit operations will become more feasible, making our design increasingly practical. Furthermore, our current implementation does not account for the impact of noise or incorporate any form of error correction or mitigation. That said, given the structural similarity of our floating-point representation to classical formats, we expect that conventional error correction techniques will be able to be adapted with minimal modification to suit our quantum framework.

\section{Conclusion}
We implemented novel quantum algorithms for efficient floating-point arithmetic and fixed-point arithmetic that that optimizes resource to be more easily implemented on quantum hardware. Our framework could be used to exploit quantum parallelism to solve optimization problems in scientific computing in general. For example, it could enable one to develop a quantum algorithm to simulate (non-linear) ordinary differential equations like the Bloch Equation for all possible physical parameter combinations at once. Such a quantum advantage could be exploited to optimize hard-to-solve cost functions, such as pulse sequence optimization in magnetic resonance imaging (MRI) \cite{jordan2021pulseseq}. Another possible application is uncertainty quantification of design parameters using Monte Carlo sampling. Essentially, any problem in scientific computing that can be easily parallelized but whose parallelization can be expensive in terms of memory would be an ideal candidate for our floating-point approach using quantum computers.

\section*{Acknowledgements}
This work was supported by the U.S. Department of Energy, Office of Science, National Quantum Information Science Research Centers, Superconducting Quantum Materials and Systems Center (SQMS), under Contract No. 89243024CSC000002. Fermilab is managed by FermiForward Discovery Group, LCC, acting under Contract No. 89243024CSC000002.
\bibliographystyle{IEEEtran}
\bibliography{main}
\end{document}